\documentclass{revtex4}
\usepackage{epsf}

\begin{document}
\title{Dineutron correlations in quasi two-dimensional systems 
 in a simplified model and possible relation to neutron-rich nuclei}

\author{Yoshiko Kanada-En'yo}
\affiliation{Yukawa Institute for Theoretical Physics, Kyoto University,
Kyoto 606-8502, Japan}

\author{Nobuo Hinohara}
\affiliation{Yukawa Institute for Theoretical Physics, Kyoto University,
Kyoto 606-8502, Japan}

\author{Tadahiro Suhara}
\affiliation{Department of Physics, Kyoto University,
Kyoto 606-8502, Japan}

\author{Peter Schuck}
\affiliation{Institut de Physique Nucl\'eaire, CNRS,
UMR 8608, Orsay, F-91505, France\\
Universit\'e Paris-Sud, Orsay, F-91505, France and\\
Laboratoire de Physique et Mod\'elisation des Milieux Condens\'es,
CNR et Universit\'e Joseph Fourier, 25 Av. des Martyrs,
BP 166, F-38042 Grenoble Cedex 9, France}

\begin{abstract}
Two-neutron correlation in the $^1S$ channel 
in quasi two-dimensional (2D) neutron systems at zero temperature
is studied by means of the BCS theory with finite-range effective nuclear
forces. The dineutron correlation in low density neutron systems 
confined in an infinite slab is investigated 
in a simplified model that neutron motion of one direction is frozen.
When the slab is thin enough,  
two neutrons form a tightly bound dineutron with a small size
in the quasi-2D system, and a Bose dineutron gas is found 
in low density limit. 
With increase of Fermi momentum, 
the neutron system changes from the Bose-gas phase to the superfluid Cooper
pair phase. The density dependence of the neutron pairing shows 
the BCS-BEC crossover phenomena at finite low-density region. 
In the transition region, the size shrinking of 
neutron pair and enhancement of pairing gap are found. 
The relation to dineutron correlation at surface of 
neutron-rich nuclei is also discussed.
\end{abstract}
\maketitle

\noindent

\section{Introduction}

Two-neutron correlations in neutron-rich nuclei 
is presently one of the fore-front subjects in the physics of unstable nuclei.
In two-neutron halo nuclei \cite{tanihata85} such as $^{11}$Li,
the dineutron correlation was theoretically 
predicted in many 
studies \cite{Hansen:1987mc,Bertsch:1991zz,Zhukov:1993aw,Esbensen:1997zz,
ikeda02,
aoyama01,Myo:2002wq,Myo:2003bh,Hagino:2005we,myo07} and is 
supported by experiments \cite{ieki93,sackett93,shimoura95,zinser97,nakamura06}.
Dineutron correlations are discussed also 
in light neutron-rich nuclei such as in $^8$He \cite{KanadaEn'yo:2007ie,
Itagaki:2008zz,Hagino:2008vm} and also 
in medium-heavy neutron-rich nuclei \cite{Matsuo:2004pr,Pillet:2007hb}
as well as asymmetric nuclear 
matter 
(for example, Refs.~\cite{bardo90,Takatsuka:1992ga,DeBlasio:1997zz,Dean:2002zx,Baldo:2005qi,Matsuo:2005vf,Margueron:2007uk,Isayev:2008wu}
and references therein).

Dineutron correlations are characterized by  strong 
spatial correlations of two neutrons in the $^1S$ channel.
Although the neutron-neutron interaction is too
weak to form a two-body bound state, the attraction in the $^1S$ channel
is rather strong as is known from the large scattering length 
$a_s=-18.5\pm 0.4$ fm \cite{teramond87}.
Originally the possible existence of a dineutron 
near the surface of nuclei
was predicted by Migdal in 1972 \cite{migdal72}. In his work, it was shown 
that two interacting particles in a potential well can form a bound state 
if there is a single-particle level with energy close to zero.
This idea describes the binding mechanism of a Borromean system
and is extended to the dineutron cluster picture and the dineutron correlation 
in two-neutron halo nuclei.

Recently, dineutron correlations 
have also been discussed from the point of view of 
Bose-Einstein condensation (BEC), which is considered along similar lines as 
deuteron  and $\alpha$-particle condensation 
suggested in dilute nuclear 
matter \cite{Takatsuka:1992ga,Bando68,Baldo:1995zz,Ropke:1998qs,Lombardo:2001ek}
as well as $\alpha$ condensate states 
in excited states of $Z=N=$ even 
nuclei \cite{Tohsaki:2001an,Yamada:2003cz,Funaki:2003af}. 
This $nn$-BEC feature
stems from the surprisingly small size of Cooper pairs recently revealed in 
the surface of nuclei \cite{Pillet:2007hb}. 
In nuclear systems, the $S=0$ $T=1$ pair correlation 
in the Bardeen-Cooper-Schrieffer (BCS) phase
has been known to be very important \cite{Ring,brink}. 
From the analysis of the spatial structure of the Cooper pair
in neutron matter, a size shrinking is found to occur
at finite low density \cite{DeBlasio:1997zz,Matsuo:2005vf,Margueron:2007uk}. 
It indicates that
the spatially correlated neutron pair may appear 
in the transition region from weak coupling BCS regime 
to strong coupling BEC regime.
Similar BCS-BEC crossover features of the spatially correlated neutron pair 
were predicted also at the surface of medium-heavy 
nuclei \cite{Matsuo:2004pr,Pillet:2007hb}. Since nuclear pairing increases as 
the size of nuclei shrinks ($\Delta \sim 12/\sqrt{A}$ MeV), one may expect that 
the BEC features are enhanced with respect to the infinite matter case.

Thus, the spatial structure of neutron pairs at the surface of neutron-rich 
nuclei attracts presently great interest. In particular, BCS-BEC cross over 
features are expected to appear in the environment of dilute neutron 
matter, such as realized 
in neutron skins \cite{Tanihata:1992wf,Suzuki:1995yc,ozawa01}. 
This has been put into 
evidence in recent studies \cite{Matsuo:2004pr,Pillet:2007hb}
where it was found that the extension 
of a $nn$-Cooper pair strongly decreases when going from inside to the 
surface before expanding again when leaving the nucleus. The minimum is 
very pronounced and the Cooper pair attains a minimum size of about 2-3fm. 
This is very small size reaching the dimension of the deuteron, that is of 
a bound state. This is also highlighted by the fact that in $^{11}$Li even 
the single 
Cooper pair is very small in the surface  \cite{Hagino:2008wt}.

The fact that $nn$-Cooper pairs resemble a bound state in the nuclear surface 
and even become appreciably smaller than in low density neutron matter 
with maximum gap value, is intriguing. In this work we want to explore the 
rather extreme hypothesis that the binding mechanism may be influenced by 
some quasi two-dimensional features. By 'quasi two-dimensional (2D)', we 
mean that the 'bound' Cooper pairs are confined within a surface layer of 
about $1-2$ fm
and that the degree of freedom in radial direction is approximately frozen. 
Such a scenario could for example be realized by the fact that the density 
distribution of valence neutrons in very neutron rich and heavy nuclei 
are in radial, say $z$-direction, concentrated in a surface layer and that 
the density distribution in $z$-direction can be approximated by a frozen 
Gaussian packet, whereas the motion of the neutrons within the 
layer is free.
Pursuing the picture to its extreme, one could imagine a slab of low 
density neutrons where in the transverse ($z$) direction 
only a single $0s$ level below the Fermi energy is active.
Such situation also could eventually be relevant in the Lasagne 
phase in neutron star crusts, 
where a low-density neutron gas may appear as a layer on the 
surface of the Lasagne
structure \cite{Ravenhall:1983uh,Lorenz:1992zz,oyamatsu93,watanabe02}. 
In any case, one could 
imagine that in the outer layer of a neutron-rich nucleus, the 
pairing properties of the valence neutrons are more or less decoupled from the 
inert core part. For example, very neutron-rich Sn isotopes with $N>82$ could represent 
such a scenario. A 'quasi-2D' slab of low density neutron matter with the 
transverse degrees of freedom approximately frozen could then, as already 
mentioned, mimic the situation. We here insist on the 2D character 
because it is well known that the dimensionality has great influence on 
bound state formation and, thus, also on the formation of tight Cooper 
pairs. Such quasi-2D paired systems with low density 
also have been investigated in cold 
atom physics e.g. hydrogen atoms on a liquid $^4$He surface
and laser cooled atoms trapped in an planar 
potential \cite{Safonov:1998zz,petrov00}.

At any rate, and for the different reasons mentioned above, it seemed to us 
interesting to study an idealized situation of quasi-2D low density 
neutron matter realized within an infinitely extended slab. The study may 
turn out academic in the end but if future more detailed and not completely 
trivial investigations will exhibit a 
link to the surprisingly strong bound state features of Cooper pairs in 
the nuclear surface, 
mentioned above, our effort will have been worth while. In this respect it 
is of interest to mention that, indeed, our study will show that in 2D matter 
a Cooper 
pair attains a minimal size of about 2 fm whereas in 3D this value is over 
twice as large.

In the present work, we investigate properties of  neutron pairs 
in quasi-2D neutron systems by using finite-range 
effective nuclear forces. 
Concerning the degree of freedom in the transverse $z$ direction
perpendicular to the 2D plane, we for simplicity assume 
a Gaussian packet as mentioned above.
We study the two-neutron wave function in this model of 
quasi-2D system and show that two neutrons form a tightly bound
state with a small size when the width $a$ of the Gaussian in $z$-direction 
is small enough.
We also investigate pairing properties of neutron matter 
based on the BCS theory. Pairing gap and size of 
the Cooper pair in quasi-2D neutron systems are analyzed, 
and  BCS-BEC crossover phenomena are discussed. 

The behaviour of $T=1$ pairing in infinite three-dimensional (3D) matter 
has been extensively investigated based on the BCS theory \cite{Bardeen:1957mv}
for symmetric matter and neutron matter
(for example, Refs.~\cite{bardo90,Takatsuka:1992ga,Dean:2002zx,Baldo:2005qi} 
and references therein).
As for a slab of nuclear matter,
properties of $T=1$ pairing in symmetric nuclear matter 
have been studied in Refs.~\cite{Baldo:1998vd,Farine:1999gs,Baldo:2003qx,
Pankratov:2008np} and it was shown that the pairing gap increases 
at the surface of the slab.
As for $T=0$ pairing, the gradual transition from BCS to BEC 
involving the deuteron formation was predicted 
in dilute symmetric matter \cite{Takatsuka:1992ga,Bando68,Baldo:1995zz,Lombardo:2001ek,Lombardo:2000tm}. 
Competition of alpha and deuteron condensation was also discussed 
\cite{Ropke:1998qs}.

There are many varieties of BCS calculations of nuclear matter
concerning, e.g., the treatment of single-particle spectra and the
pairing interaction. For convenience, effective nuclear 
forces have been used in some calculations on pairing in nuclear matter 
instead of dealing with realistic bare nuclear forces.
It was shown that the Gogny force \cite{gognyd1,gognyd1s}, a
finite-range effective nuclear force, 
gives similar behaviour of nuclear matter pairing 
to that calculated from realistic nuclear forces with the Brueckner-HF (BHF) 
approximation, as well as reasonable pairing properties 
in finite nuclei
\cite{DeBlasio:1997zz,gognyd1,gognyd1s,kucharek89,Garrido:1999at}. 
Therefore, we adopt the Gogny D1S force in the present BCS calculations.
We also use the Minnesota force \cite{minnesota} which reproduces 
features of $S$-wave $N$-$N$ scattering in $T=0$ and $T=1$ channels, and
nucleus-nucleus scattering in the light nuclear region 
such as the $\alpha$-$\alpha$ system, as well as properties of light nuclei.
The Minnesota force is often used in structure studies 
of light neutron-rich nuclei.

This paper is organized as follows. 
In the next section, we explain our simplified model of
quasi-2D systems and formulation of the present work.
The effective nuclear forces are described in section \ref{sec:interaction}.
In section \ref{sec:results}, the results obtained for 
a two-neutron system and neutron matter are shown, and in section 
\ref{sec:discussion} discussions are outlined.
Finally, we give a summary and outlook in section \ref{sec:summary}.

\section{Formulation of Nuclear 2D Pairing} \label{sec:formulation}

Here we explain our framework of the simplified model 
of quasi-2D neutron system confined in a slab with a certain thickness.
We first describe the wave function and the Hamiltonian for a
quasi-2D neutron matter system. We then explain the isolated two-neutron 
system 
and give the formalism for the quasi-2D system within the BCS theory.

\subsection{Quasi-2D neutron systems}

According to the outline of the Introduction, 
we propose a model for a system of neutrons 
confined in a slab with a certain thickness. 
We assume that the neutrons in transverse ($z$) direction are represented by a 
simple Gaussian packet of size $a$.
As already mentioned, this size may mimic the concentration 
of the  amplitudes of single particle wave functions 
in a surface layer, where  $z$ corresponds to the radial direction
in finite neutron-rich nuclei.
 
Therefore, the wave function for a single neutron is written as 
\begin{equation}
\Phi({\bf r})=\Phi^{2D}({\bf r}_\perp,\chi)\otimes \phi^{0s}(z),
\end{equation}
\begin{equation}
\phi^{0s}(z)=\left( \frac{1}{\pi a^2}\right)^{1/4} 
\exp\left[-\frac{z^2}{2a^2}\right],
\end{equation}
where ${\bf r}_\perp$ indicates the coordinates in the slab, $(x,y)$,
and $\chi$ is the intrinsic spin.
The thickness, i.e., the rms width, of the slab is 
$2\sqrt{\langle z^2 \rangle}=\sqrt{2}a$.
In a similar way, 
a $N$-neutron wave function is given as
\begin{equation}
\Phi({\bf r_1},\cdots,{\bf r_N},\chi_1,\cdots,\chi_N)=
\Phi^{2D}({\bf r}_{1\perp},\cdots,
{\bf r}_{N\perp},\chi_1,\cdots,\chi_N)
\otimes \phi^{0s}(z_1)\cdots\phi^{0s}(z_N).
\end{equation}
Here $\Phi$ and $\Phi^{2D}$ are the antisymmetrized many-body wave functions.

Neutrons are interacting with each other via two-body nuclear forces.
The interaction $V (r_{ij})$ 
between two neutrons 
is the two-body effective nuclear force in the $T=1$ channel.
Here $r_{ij}=|{\bf r_i}-{\bf r_j}|$. 
Because the $S$-wave neutron pair correlation is studied in the present work,
only the central part of the effective nuclear force is considered. 
The Gogny and Minnesota forces are represented by a superposition of two 
Gaussians,
\begin{equation}
V(r)=\sum_{m=1}^2 \left(W_m+B_mP_\sigma-H_mP_\tau-M_mP_\sigma P_\tau
\right)\exp\left[ -\frac{r^2}{b_m^2} \right].\label{eq:gogny}
\end{equation}

We freeze the the degree of freedom in transverse ($z$) direction,
as described in the Introduction. By integrating over $z_i$ coordinates
the three-dimensional Schr\"odinger equation 
for $\Phi$ is reduced to the two-dimensional equation 
for $\Phi^{2D}$ with respect to ${\bf r}_{i\perp}$. 
 Then we get the following equation for
quasi-2D neutron systems, 
\begin{equation}
H^{2D}\Phi^{2D}({\bf r}_{1\perp},\cdots,{\bf r}_{N\perp},\chi_1,\cdots,\chi_N)
=E^{2D}\Phi^{2D}({\bf r}_{1\perp},\cdots,
{\bf r}_{N\perp},\chi_1,\cdots,\chi_N),\label{eq:schrodinger2D}
\end{equation}
\begin{equation}
H^{2D}=\sum_i t_{i\perp} +\sum_{i<j} V^{2D} (r_{ij\perp}),
\end{equation}
where the 2D-kinetic term and the 2D-interaction term are
\begin{equation}
t_{\perp}=
-\frac{\hbar^2}{2m}\frac{\partial^2}{\partial {\bf r}_{\perp}^2}
=-\frac{\hbar^2}{2m}\left(\frac{\partial^2}{\partial x^2}
+\frac{\partial^2}{\partial y^2}\right),
\end{equation}
\begin{eqnarray}\label{eq:v2d}
V^{2D} (r_\perp)&=&
\left\langle\phi^{0s}(z_1)\phi^{0s}(z_2)\right|
V (r)
\left|\phi^{0s}(z_1)\phi^{0s}(z_2)\right\rangle \nonumber\\
& =& 
\sum_{m=1}^2 Z_{b_m}(a)
\left(W_m+B_mP_\sigma-H_mP_\tau-M_mP_\sigma P_\tau\right)
\exp\left[ -\frac{r_\perp^2}{b_m^2} \right],\\
Z_{b_m}(a)&\equiv&\left(\frac{b_m^2}{b_m^2+2a^2}\right)^{1/2}
\label{eq:strength2D}.
\end{eqnarray}

As seen from (\ref{eq:v2d}) and (\ref{eq:strength2D}), 
the strength of the two-body interaction $V^{2D}(r_\perp)$ 
in the quasi-2D system depends on the thickness of the slab
with the factor $Z_{b_m}(a)$.
The factor decreases  
with the increase of the width parameter $a$.
The thickness 
of the slab is only reflected in a weakening of the interaction 
for bigger slab sizes in our genuinely two-dimensional model.
These are, of course, strongly simplified, may be oversimplified, assumptions 
to describe 
the detailed behaviour of 
the valence neutrons because the Pauli effect from the core should be
more complicated and  
a long tail of weakly bound neutrons may be not so simple.
However, it is expected that 
such effects may effectively be taken into account 
by a modification of the width parameter $a$. 
In the present work, we analyze 
properties of neutron pairs 
in quasi-2D neutron systems 
for various values of the slab size $a$.

\subsection{The isolated two-neutron system}
For later comparison we consider 
the $S$-wave relative wave function $\Phi^{nn}(r_\perp,\chi_1,\chi_2)$ 
for an isolated  two-neutron system in the quasi-2D space. It is obtained 
by solving 
the Schr\"odinger equation, 
\begin{eqnarray}\label{eq:schro-nn}
&&\left(-\frac{\hbar^2}{m}\left(
\frac{\partial^2}{\partial {r}_\perp}
+\frac{1}{r_\perp}\frac{\partial}{\partial r_\perp}\right)
+V^{2D} (r_\perp)\right)\Phi^{nn}(r_\perp,\chi_1,\chi_2)=
E^{2D}_{nn}\Phi^{nn}(r_\perp,\chi_1,\chi_2),\\
&&\Phi^{nn}=\psi^{nn}(r_\perp)\otimes X_{S=0}(\chi_1,\chi_2).
\end{eqnarray}
 
$\psi^{nn}(r_\perp)$ is the spatial wave function 
normalized as $\int d^2r_\perp |\psi^{nn}(r_\perp)|^2=1$,
and $X_{S=0}$ is the intrinsic-spin wave function for the spin-zero channel. 

It is important to note that in two dimensions, a bound state can form at 
any arbitrarily small 
attraction \cite{landau89}.
Since the nuclear force in the $^1S$ channel
is attractive at low energy, a two-neutron bound state in the quasi-2D space 
is obtained with an arbitrary width $a$.
We calculate the exact solution of the bound state with a Gaussian 
expansion method. 
The size of the two-neutron bound state is given as
\begin{equation}
\xi_\perp^{nn}=\sqrt{\int d^2r_\perp r^2_\perp|\psi^{nn}(r_\perp)|^2.}
\label{eq:xinn}
\end{equation}
Please note that 
the size in 3D is given as $\xi=\sqrt{(\xi^{nn}_\perp)^2+a^2}$.

\subsection{Quasi two-dimensional neutron matter within BCS theory}

In this section,
we investigate the behaviour of neutron pairs 
in quasi-2D infinite neutron matter in the slab 
by means of the BCS theory,
which is equivalent to the HFB approximation 
in a homogeneous case \cite{Ring,brink}. 
The application of BCS theory with the Gogny force 
to a slab of nuclear matter 
was, e.g., already performed in Ref.~\cite{Farine:1999gs}.

Let us explain the equations for $^1S$ pairing in quasi-2D neutron matter.
We apply the BCS theory to the 2D Schr\"odinger equation 
(\ref{eq:schrodinger2D}).
The $S$-wave pairing gap depends on $|{\bf p}_\perp|$ and it 
is written as 
\begin{equation}\label{eq:2dgap}
\Delta(p)=-\frac{1}{2}\int \frac{d^2k_\perp}{(2\pi)^2}v_{\rm pp}
({\bf p}_\perp-{\bf k}_\perp)
\frac{\Delta(k)}{\sqrt{(e(k)-\mu)^2+\Delta(k)^2}},
\end{equation}
\begin{equation}
 e(p)=\frac{{\bf p}^2_\perp}{2m}+V^{HF}(p).\label{eq:single-ene}
\end{equation}
Here $p$ and $k$ denote $|{\bf p}_\perp|$ and  $|{\bf k}_\perp|$, respectively, 
and $v_{\rm pp}$ is the pairing force,
\begin{equation}
v_{\rm pp}({\bf p}_\perp-{\bf k}_\perp)= \sum^2_{m=1}
Z_{b_m}(a)\left(W_m-B_m-H_m+M_m\right)
\left(\sqrt{\pi}b_m\right)^2 \exp\left[-\frac{1}{4}b_m^2
({\bf p}_\perp-{\bf k}_\perp)^2 \right].
\end{equation}
$V^{HF}(p)$ is the Hartree-Fock potential, 
\begin{equation}
 V^{HF}(p)=\frac{\rho}{2}
\sum^2_{m=1}Z_{b_m}(a)\left(2W_m+B_m-2H_m-M_m\right)
\left(\sqrt{\pi}b_m\right)^2 
-\int \frac{d^2k_\perp}{(2\pi)^2}v_{\rm ph}({\bf p}_\perp-{\bf k}_\perp) v^2_k,
\end{equation}
where the exchange term is given as
\begin{equation}
v_{\rm ph}({\bf p}_\perp-{\bf k}_\perp)= \sum^2_{m=1}Z_{b_m}(a)\left(W_m+2B_m-H_m-2M_m\right)
\left(\sqrt{\pi}b_m\right)^2 \exp\left[-\frac{1}{4}b_m^2
({\bf p}_\perp-{\bf k}_\perp)^2 \right].
\end{equation}
The occupation probability is
\begin{equation}
v_k^2=\frac{1}{2}\left(1-\frac{e(k)-\mu}{\sqrt{(e(k)-\mu)^2+\Delta(k)^2}}\right),
\end{equation}
and the chemical potential $\mu$ is determined so as to satisfy 
the number equation, 
\begin{equation}
\frac{\rho}{2}=\frac{k^2_F}{4\pi}=\int \frac{d^2k_\perp}{(2\pi)^2}v^2_{k},
\end{equation}
where $\rho$ and $k_F$ are the density and the Fermi momentum 
in 2D, respectively.
In this paper, the $k_F$ dependence of pairing properties in neutron matter
is discussed as a function of the ratio $k_F/k_0$,  
where $k_0=1.36$ fm$^{-1}$ is the Fermi momentum at
the normal density of 3D symmetric nuclear matter.

The energy per nucleon for the BCS state is 
\begin{equation}
E_{BCS}\equiv \frac{1}{\rho}
\langle \Phi_{\rm BCS}|H^{2D}|\Phi_{\rm BCS} \rangle
=\frac{1}{\rho}\int \frac{d^2k_\perp}{(2\pi)^2} 
\left[\frac{{\bf p}^2_\perp}{2m}+\frac{1}{2}V^{HF}(k) \right]2v^2_{k}+E_{pair},
\end{equation}
where the integrated pair energy $E_{pair}$ is
\begin{equation}
E_{pair}\equiv-\frac{1}{\rho}\int \frac{d^2k_\perp}{(2\pi)^2} \Delta (k) \kappa(k),
\end{equation}
with 
\begin{equation}
\kappa(k)= u_kv_k=\frac{1}{2}\frac{\Delta(k)}{\sqrt{(e(k)-\mu)^2+\Delta(k)^2}}.
\end{equation}
The energy per nucleon for the HF state $\Phi_{\rm HF}$ is determined from
\begin{equation}
E_{HF}\equiv \frac{1}{\rho}\langle \Phi_{\rm HF}|H^{2D}|\Phi_{\rm HF} \rangle.
\end{equation}

To analyze the spatial structure of a $nn$-Cooper pair
it is useful to study the pair wave function in the coordinate space. 
It is defined by the Fourier transform of the
anomalous density $\kappa(k)$,
\begin{equation}
\Psi_{\rm pair}(r_\perp)\equiv n_0 \langle \Phi_{\rm BCS}|
a^\dagger({\bf r}'_{\perp} \uparrow) a^\dagger({\bf r}''_\perp \downarrow) 
|\Phi_{\rm BCS} \rangle
=n_0\int \frac{d^2k_\perp}{(2\pi)^2} \kappa(k) e^{i{\bf k}_\perp \cdot 
({\bf r}'_\perp-{\bf r}''_\perp)}.
\end{equation}
Here $r_\perp=|{\bf r}'_\perp-{\bf r}''_\perp|$, and
$n_0$ is the normalization factor so that 
$\int d^2r_\perp |\Psi_{\rm pair}(r_\perp)|^2=1$. 
The pair size (coherence length) $\xi_\perp$ is calculated 
from the root-mean-square distance of the pair wave function
\begin{equation}
\xi_\perp\equiv \sqrt{\langle r^2_\perp\rangle},
\end{equation}
where
\begin{equation}
\langle r^2_\perp\rangle=
\int d^2r_\perp r^2_\perp|\Psi_{\rm pair}(r_\perp)|^2. 
\end{equation}
In the low-density limit $k_f\rightarrow 0$,
the pair wave function $\Psi_{\rm pair}$ and the energy 
$2E_{BCS}$ approach $\psi^{nn}$ and $E^{2D}_{nn}$
of the two-neutron bound state of section II.B.

\section{Effective nuclear forces 
and pairing properties in 3D neutron matter}
\label{sec:interaction}

As already mentioned, in this work we use the Gogny D1S force \cite{gognyd1,
gognyd1s} and the Minnesota force \cite{minnesota}.
In the $T=1$ channel, the Gogny force is density 
independent and is represented by a superposition of two 
Gaussians (Eq.~(\ref{eq:gogny})). 
The parameters for the Gogny D1S force are,
\begin{eqnarray}
&& b_1=0.7 {\rm fm},\qquad b_2= 1.2 {\rm fm} \nonumber\\
&& W_1=-1720.30 {\rm MeV},\qquad B_1=1300.00 {\rm MeV} \nonumber\\
&& H_1=-1813.53 {\rm MeV},\qquad M_1=1397.60 {\rm MeV} \nonumber\\
&& W_2=103.64 {\rm MeV},\qquad B_2=-163.48 {\rm MeV} \nonumber\\
&& H_2=162.81 {\rm MeV},\qquad M_2=-223.93 {\rm MeV}. 
\end{eqnarray}
The Minnesota force is also given by a sum of finite-range Gaussians, 
\begin{eqnarray}
V(r)&=&
\frac{1}{2}(1-P_\sigma)\frac{1}{2}(1+uP_r)
\left(V_R\exp(-\kappa_R r^2)+V_s\exp(-\kappa_s r^2)\right)
\nonumber\\
&&+\frac{1}{2}(1+P_\sigma)\frac{1}{2}(1+uP_r)
\left(V_R\exp(-\kappa_R r^2)+V_t\exp(-\kappa_t r^2)\right),\\
V_R&=&200.0 {\rm MeV}, \kappa_R=1.487 {\rm fm}^{-2},\nonumber\\
V_s&=&-91.85 {\rm MeV}, \kappa_s=0.465 {\rm fm}^{-2},\nonumber\\
V_t&=&-178.0 {\rm MeV}, \kappa_t=0.639 {\rm fm}^{-2}.
\end{eqnarray}
We adopt for the 
exchange-mixture parameter the value of $u=1.0$.
Then the first (second) term gives the two-body force in the 
$S=0$ $T=1$ ($S=1$ $T=0$) channel.

In principle the Gogny force represents an effective in medium force and the
Minnesota force a phenomenological nuclear force in free space.
It has, however, been shown that the Gogny D1S force 
gives similar behaviour for the $^1S$ pairing gap
of symmetric nuclear matter as that calculated from realistic nuclear forces
with the BHF+BCS approximation, and 
also describes pairing properties 
in finite nuclei \cite{DeBlasio:1997zz,gognyd1,gognyd1s,kucharek89,
Garrido:1999at}. 
The Minnesota force reasonably reproduces 
the $S$-wave $N$-$N$ scattering in the $S=0$  $T=1$ channel as well as 
the $S=1$  $T=0$ channel. It also well describes 
nucleus-nucleus scattering such as the
$\alpha$-$\alpha$ system, as well as properties of light nuclei. 
Because the Minnesota force 
has no density-dependent term in the $S=1$ $T=0$ channel, it 
can not describe the saturation properties of symmetric 
nuclear matter. However, it gives a similar density dependence of 
neutron matter energy as the one obtained with the Gogny D1S force,
as shown later.

\begin{figure}[th]
\epsfxsize=7 cm
\centerline{\epsffile{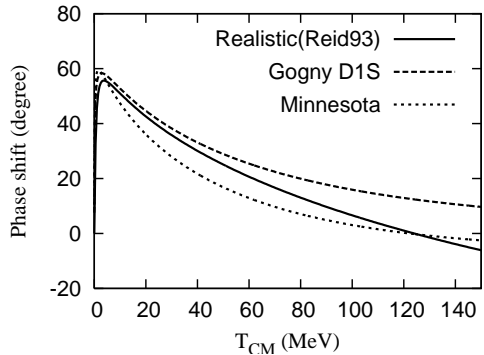}}
\vspace*{8pt}
\caption{Phase shift of $N$-$N$ scattering in the $^1S$ channel
obtained with the Gogny D1S force and Minnesota force. 
$^1S$ phase shift of a realistic force, 
Reid93 potential \cite{Stoks:1994wp}, 
which fits the experimental proton-proton scattering
cross sections, is also shown.
\label{fig:phase}
}
\end{figure}

The values of the $n$-$n$ scattering length $a_s$ 
in the $^1S$ channel obtained with the Gogny D1S force
and Minnesota force 
are $a_s=-12.1$ fm and $a_s=-16.8$ fm, respectively.
These values semiquantitatively agree with 
the experimental value $a_s=-18.5\pm 0.4$ fm \cite{teramond87}.
In Fig.~\ref{fig:phase}, the calculated 
phase shifts of $N$-$N$ scattering in the $^1S$
channel are shown as well as those obtained with 
a realistic force. Both the Gogny D1S and Minnesota forces 
show a reasonable reproduction of the $^1S$-wave scattering.

\begin{figure}[th]
\epsfxsize=6 cm
\centerline{\epsffile{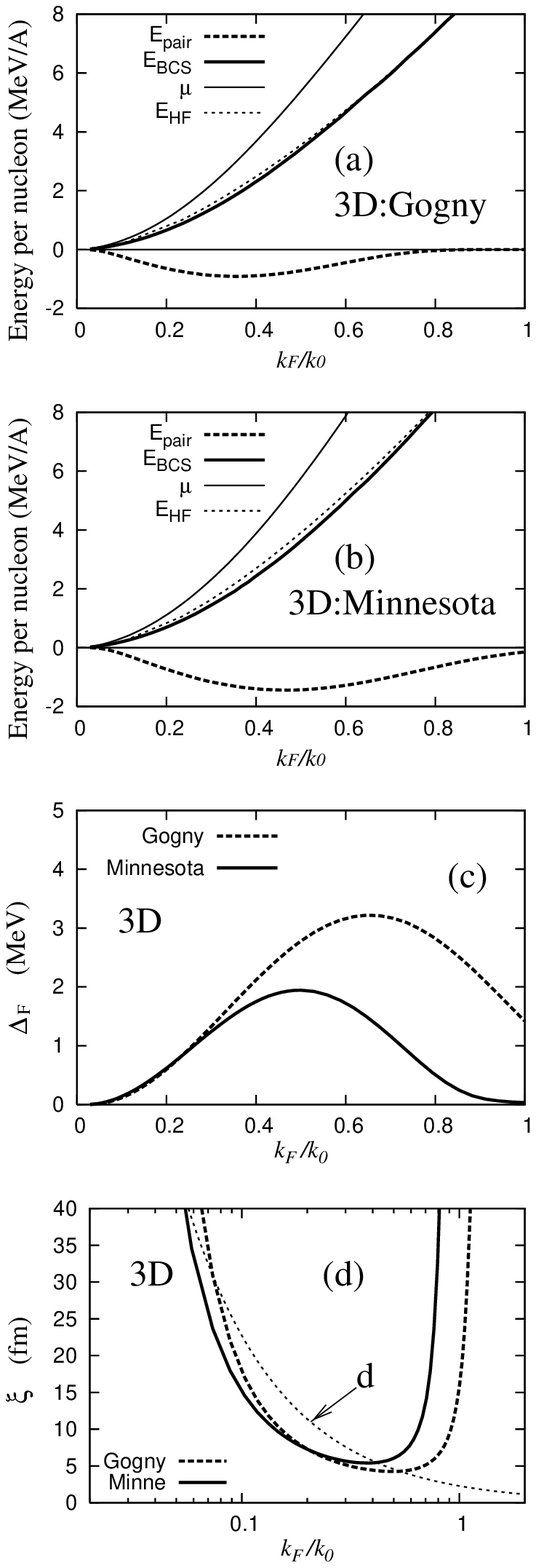}}
\vspace*{8pt}
\caption{Energy, pairing gap, chemical potential and pair size in 
3D neutron matter as a function of $k_F/k_0$ ($k_0=1.36$ fm$^{-1}$) 
obtained with the Gogny D1S force and Minnesota force.
(a)(b) Energy ($E_{BCS}$), 
the integrated pair energy ($E_{pair}$) per nucleon 
and the chemical potential ($\mu$)
of the BCS state, and the energy of the HF state ($E_{HF}$).
(c) Gap energy at the Fermi momentum $\Delta_F=\Delta(k=k_F)$.
(d) Pair size $\xi$.
The average inter-neutron distance $d=\rho^{-1/3}$ is shown by a thin 
dashed line. 
\label{fig:fig3d}
}
\end{figure}

Figure \ref{fig:fig3d} shows the properties of infinite 3D neutron matter
calculated with the Gogny D1S and Minnesota forces 
by means of the BCS theory as well as from the HF calculations.
Figures \ref{fig:fig3d}(a) and \ref{fig:fig3d}(b) show the 
the energy ($E_{BCS}$) and  
the integrated pair energy ($E_{pair}$)  per nucleon, and
the chemical potential ($\mu$) of the BCS state
are shown as well as 
the energy of the HF state ($E_{HF}$) for 3D neutron matter 
with the Gogny D1S and Minnesota forces, 
respectively. 
The density dependence of the neutron matter 
energy obtained with both forces is in reasonable agreement with
that calculated with ab initio methods
using realistic nuclear forces \cite{Akmal:1998cf,Carlson:2003wm}.

In Fig.~\ref{fig:fig3d}(c), the 
pairing gap of neutron matter is shown. 
The difference between the Gogny D1S and Minnesota forces is not 
large in the low-density region $k_F/k_0 < 0.4$.
In the $k_F/k_0 \ge 0.4$ region, the Minnesota force gives a  much 
smaller pairing gap than the Gogny D1S force because it has a larger 
repulsive core
with 108.15 MeV hight at $r=0$. 

Let us make a comment on the pairing gap in 3D neutron matter investigated 
in different approaches.
There are many calculations of the pairing gap in nuclear matter 
using realistic nuclear forces (for reviews \cite{Takatsuka:1992ga,Dean:2002zx}). 
In the standard BHF+BCS approximation, 
the pairing gap in 3D neutron matter is about 3 MeV at the peak position 
$k_f/k_0\sim 0.6$ (figure 12 of Ref.~\cite{Dean:2002zx}), which is
consistent with the pairing gap calculated in BCS theory
with the Gogny force.
In  different approximations with 
medium polarization effects, a quenching of the pairing gap 
in neutron matter was suggested
\cite{Margueron:2007uk,schulze96,Lombardo:2001vp,Shen:2002pm,lombardo04,Cao:2006gq}.

The size $\xi$ of a Cooper pair in 3D neutron matter is
plotted as a function of $k_F/k_0$ in Fig.~\ref{fig:fig3d}(d).
As is already  
discussed in Refs.~\cite{DeBlasio:1997zz,Matsuo:2005vf,Margueron:2007uk}, 
the pair size shrinking in 3D neutron matter is seen at finite low $k_F$. 
Similarly to the pairing gap, 
the difference of the pair size between 
the Gogny and Minnesota forces is small in the low-density region.
In the $k_F/k_0\ge 0.4$ region,
the pair size increases more rapidly because of the weaker pairing
with the Minnesota force than is the case with the Gogny D1S force.
As we show later, also in quasi-2D neutron matter
these two forces give qualitatively similar pairing behaviour 
for low $k_F$. 
With both  forces,
the $\xi< d$ region ($d$ is the average inter-neutron 
distance $d=\rho^{-1/3}$) appears at finite low $k_F$, i.e., at low density.
It shows the BCS-BEC crossover feature 
as suggested by Matsuo \cite{Matsuo:2005vf}.
 
\section{Results}\label{sec:results}

In this section,  two-neutron correlations in quasi-2D neutron systems are 
investigated by performing BCS calculations for 2D neutron matter 
with the Gogny D1S  and Minnesota forces.
The results for various values of $a$ of the confinement in the frozen 
direction ($z$) are analyzed, and the behaviour of the Cooper pair 
in quasi-2D neutron matter is discussed as a function of $k_F/k_0$.

\subsection{Two-neutron bound state in quasi-2D system}

By solving the Schr\"odinger equation (\ref{eq:schro-nn}),
the
 wave function $\psi^{nn}(r_\perp)$ of two isolated neutrons in a
quasi-2D system is obtained.
While a two-body bound state in 3D forms only at sufficiently strong
attraction,
a bound state in 2D can form, as already mentioned, at any arbitrarily small 
attraction. Since the $^1S$ nuclear force 
is attractive at low energy, the two-neutron bound state is formed 
in quasi-2D systems with arbitrary width $a$.

The energy $E_{nn}^{2D}$ and the pair size
$\xi_\perp^{nn}$ of the bound state for $a=1$, 
$a=2$ and $a=4$ fm are given in Table \ref{tab:rmsr-2n}.
Energy and size of the dineutron 
decrease as the width $a$ becomes small. 
In the quasi-2D system with a small width of $a=1$ fm,  
the dineutron size is as small as that of the 
deuteron  in 3D.
On the other hand, in the quasi-2D system with a width $a=4$ fm, the
dineutron has a small binding energy, less than 0.1 MeV, 
and a size larger than 20 fm.
We, therefore, see that in our model the width of the slab has a strong 
influence on the two body correlations.  

It is very interesting that, although two neutrons 
are unbound in  3D,
they form a tightly bound dineutron state with a size comparable 
to deuteron dimensions when they are confined in 
a sufficiently thin slab.
To form the small-sized dineutron, 
the reduction of the dimension
from three to two is crucial.
In the present model, neutrons are confined in a Gaussian packet 
with respect to the frozen $z$-direction.

In slab direction,
the potential part $V^{2D}(r_\perp)$ is given by almost the same form 
as the original effective nuclear force $V(r)$ except for 
the factor 
$Z_{b_m}(a)$ in (\ref{eq:v2d}).
This factor is smaller than one, depending on $a$. For example, 
for the case of $a=1$ fm and 
the interaction range $b_2=1.2$ fm as with the long-range attractive term
of the Gogny D1S force, one gets $Z_{b_m}(a)=0.59$, that is 
the strength of $V^{2D}(r_\perp)$ is smaller
than that in the original interaction $V(r)$.
Let us remind the reader that the size of the dineutron state
is determined by the balance of the kinetic energy
and the potential energy. In a 2D system, the kinetic 
effect is weaker and a small-sized two-particle bound 
state can more easily be formed than in the 3D case.

\begin{figure}[th]
\epsfxsize=10 cm
\centerline{\epsffile{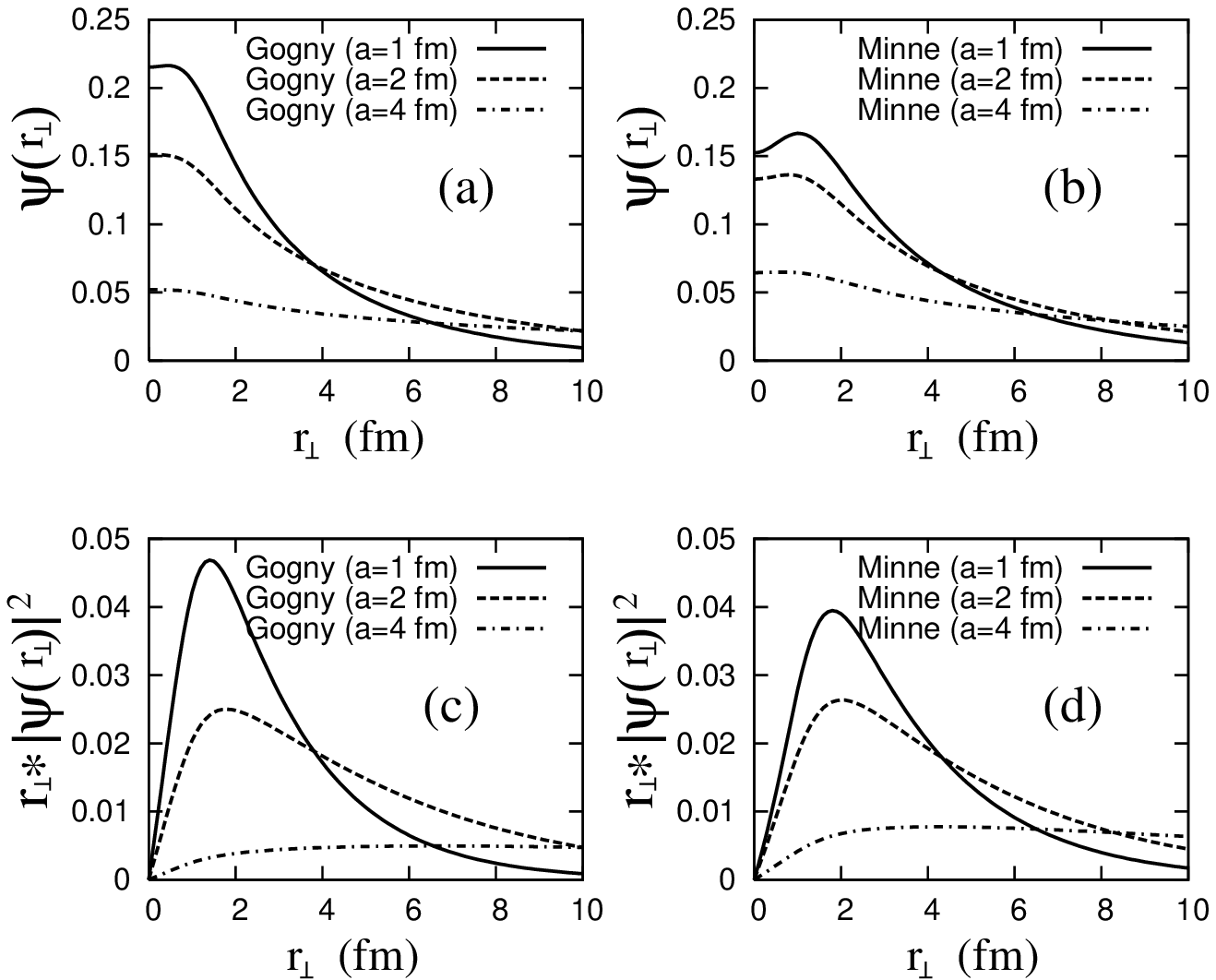}}
\vspace*{8pt}
\caption{Two-neutron wave function $\psi^{nn}(r_\perp)$ 
in quasi-2D systems with the width parameter
$a=1,2,$ and 4 fm calculated with (a) the Gogny D1S force and
(b) Minnesota force. (c)(d) Same as (a)(b) but for the
probability density $r_\perp |\psi^{nn}(r_\perp)|^2$. 
\label{fig:wf-2n}
}
\end{figure}

In Fig.~\ref{fig:wf-2n}, 
the wave function $\psi^{nn}(r_\perp)$ 
and the probability density $r_\perp |\psi^{nn}(r_\perp)|^2$ of two isolated 
neutrons 
in a quasi-2D system are shown as a function of the relative
coordinate $r_\perp$.
In the quasi-2D system with $a=1$ and $a=2$ fm,  
the probability density has a peak around $r_\perp=2$ fm,
which shows the significant spatial correlation between the
neutrons. In a quasi-2D system with $a=4$ fm,
the density distributes in a broad region, and the spatial
correlation is much weaker.

\begin{table}[ht]
\caption{ 
\label{tab:rmsr-2n} Energy $E^{2D}_{nn}$ 
and size $\xi^{nn}_\perp$ of the
two-neutron bound state $\psi^{nn}(r_\perp)$ 
in the quasi-2D system with the width parameter 
$a=1,2,$ and 4 fm. 
The energy and the size 
$\xi^{pn}$ for a deuteron in 3D space
obtained with the Minnesota force is also
shown for comparison.} 
\begin{center}
\begin{tabular}{ccccc}
\hline
&\multicolumn{3}{c}{Gogny D1S}\\
 & $a=1$ & $a=2$ & $a=4$ \\ 
$E^{2D}_{nn}$ (MeV) & $-$2.7 &	$-$0.64 &	$-$0.03 \\
$\xi^{nn}_\perp$ (fm)& 3.5 & 	6.8 & 	33 \\
\hline
&\multicolumn{3}{c}{Minnesota}\\
 & $a=1$  & $a=2$ & $a=4$ \\ 
$E^{2D}_{nn}$ (MeV)& $-$1.9 &	$-$0.74 &	$-$0.06 \\
$\xi^{nn}_\perp$ (fm) & 4.3 &	6.5 &	21 \\
\hline
&\multicolumn{4}{c}{Deuteron with Minnesota}\\
$E$ (MeV)  &\multicolumn{3}{c}{$-$2.2}\\
$\xi^{pn}$ (fm)  &\multicolumn{3}{c}{3.9}\\
\hline
\end{tabular}
\end{center}
\end{table}

\subsection{Quasi-2D neutron matter}

We now extend  
the BCS calculations to quasi-2D neutron matter 
with  varying width parameters $a$.

\begin{figure}[th]
\epsfxsize=6 cm
\centerline{\epsffile{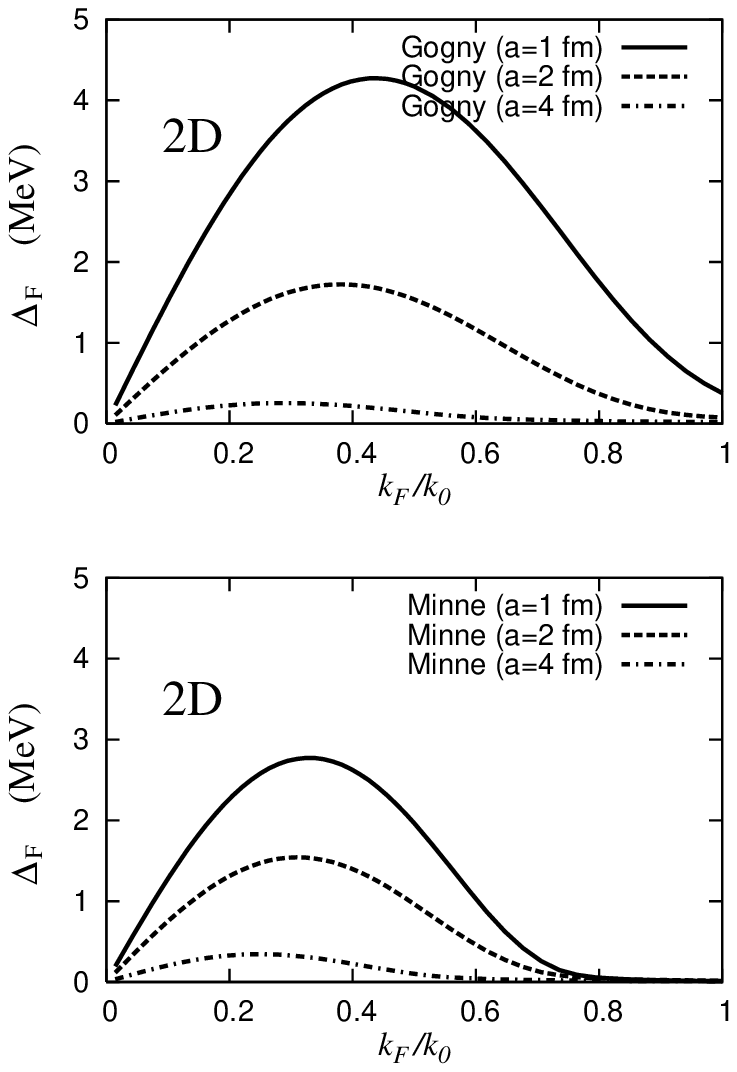}}
\vspace*{8pt}
\caption{Pairing gap at the Fermi momentum $\Delta_F=\Delta(k=k_F)$
in quasi-2D neutron matter as a function of $k_F/k_0$
($k_0=1.36$ fm$^{-1}$). 
Upper and lower panels show
the results with the Gogny D1S force and Minnesota force,
respectively.
\label{fig:gap-linear}
}
\end{figure}

\begin{figure}[th]
\epsfxsize=10 cm
\centerline{\epsffile{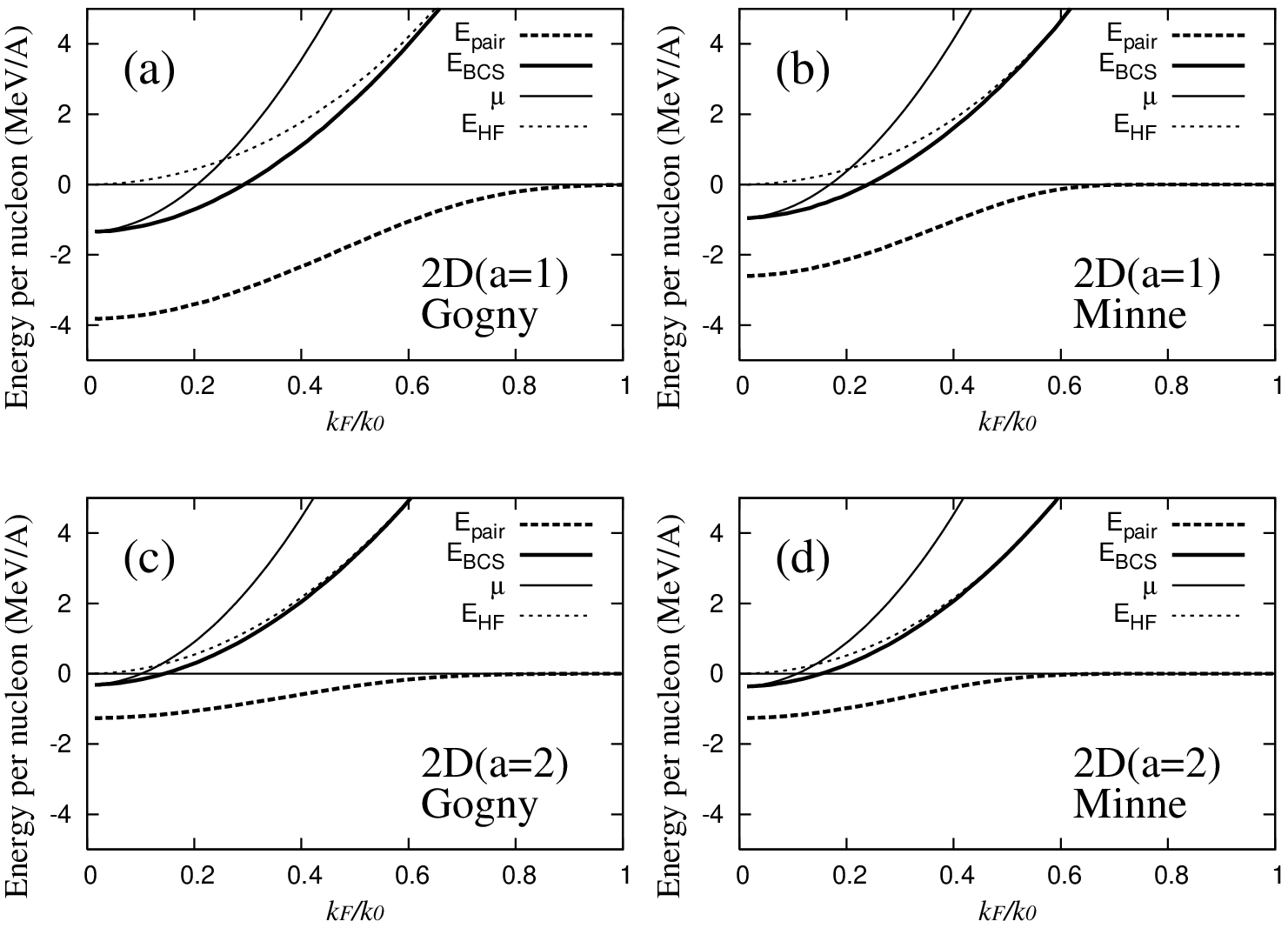}}
\vspace*{8pt}
\caption{\label{fig:hfene}
Energy ($E_{BCS}$), the integrated pair energy ($E_{pair}$) per nucleon 
and the chemical potential ($\mu$) of the BCS state, and energy  
of the HF state ($E_{HF}$) as a function of $k_F/k_0$
($k_0=1.36$ fm$^{-1}$). 
(a) and (b): Energy of the quasi-2D neutron matter with $a=1$ fm 
obtained with the Gogny D1S force and Minnesota force.
(c) and (d): Energy of the quasi-2D neutron matter with $a=2$ fm. 
}
\end{figure}

The pairing gap at the Fermi momentum $\Delta_F\equiv\Delta(k=k_F)$ 
as a function of $k_F/k_0$ is shown in Fig.~\ref{fig:gap-linear}.
The Gogny D1S and Minnesota forces give qualitatively similar results
in the low $k_F$ region. 
At $k_F/k_0 \ge 0.3$,
the pairing gap is more suppressed in case of the Minnesota force,
because, as already mentioned, the short-range repulsion is larger than with 
the Gogny force.
For $a=1$ fm, the pairing gap obtained with the Gogny D1S force
has a peak  about 4 MeV high 
at $k_F/k_0 \sim 0.4$, 
while that calculated with the Minnesota force 
has a peak  $\sim 3$ MeV high at $k_F/k_0 \sim 0.3$. 
With the increase of the width $a$ of the slab, the pairing gap 
$\Delta_F$ is quenched because the attraction of the pairing force
is  weaker in a wider slab.

In Fig.~\ref{fig:hfene}, the energy ($E_{BCS}$) and  
the integrated pair energy ($E_{pair}$)  per nucleon, and
the chemical potential ($\mu$) of the BCS state
are shown as well as 
the energy of the HF state ($E_{HF}$) calculated without pairing.
Compared with the energy of 3D neutron matter shown in 
Fig.~\ref{fig:fig3d}, 
one of the striking features of quasi-2D neutron matter is that
$E_{BCS}$, $E_{pair}$, and the chemical potential have 
finite negative values even in $k_F\rightarrow 0$ limit.
This is because, in the $k_F\rightarrow 0$ limit, 
the system goes to the strong coupling BEC limit, where
Cooper pairs become identical with the two-particle bound state. 
In this limit, $-2E_{BCS}$ and $-2\mu$ become equal to
the dineutron binding energy. As soon as the chemical potential turns negative, 
there is true binding between the two neutrons and the strong coupling BEC 
feature is born out. Therefore, in quasi-2D neutron matter one can, contrary 
to the 3D case, follow the transition from BEC to the BCS regime. On the other 
hand,
because of the same reason, 
the pairing gap $\Delta_F$ in the 
quasi-2D neutron matter shows a linear behaviour at the  low $k_F$ 
limit (Fig.~\ref{fig:gap-linear}), 
while the pairing gap disappears with a horizontal tangent at small $k_F$ 
in 3D neutron matter (Fig.~\ref{fig:fig3d}(c)).

\begin{figure}[th]
\epsfxsize=10 cm
\centerline{\epsffile{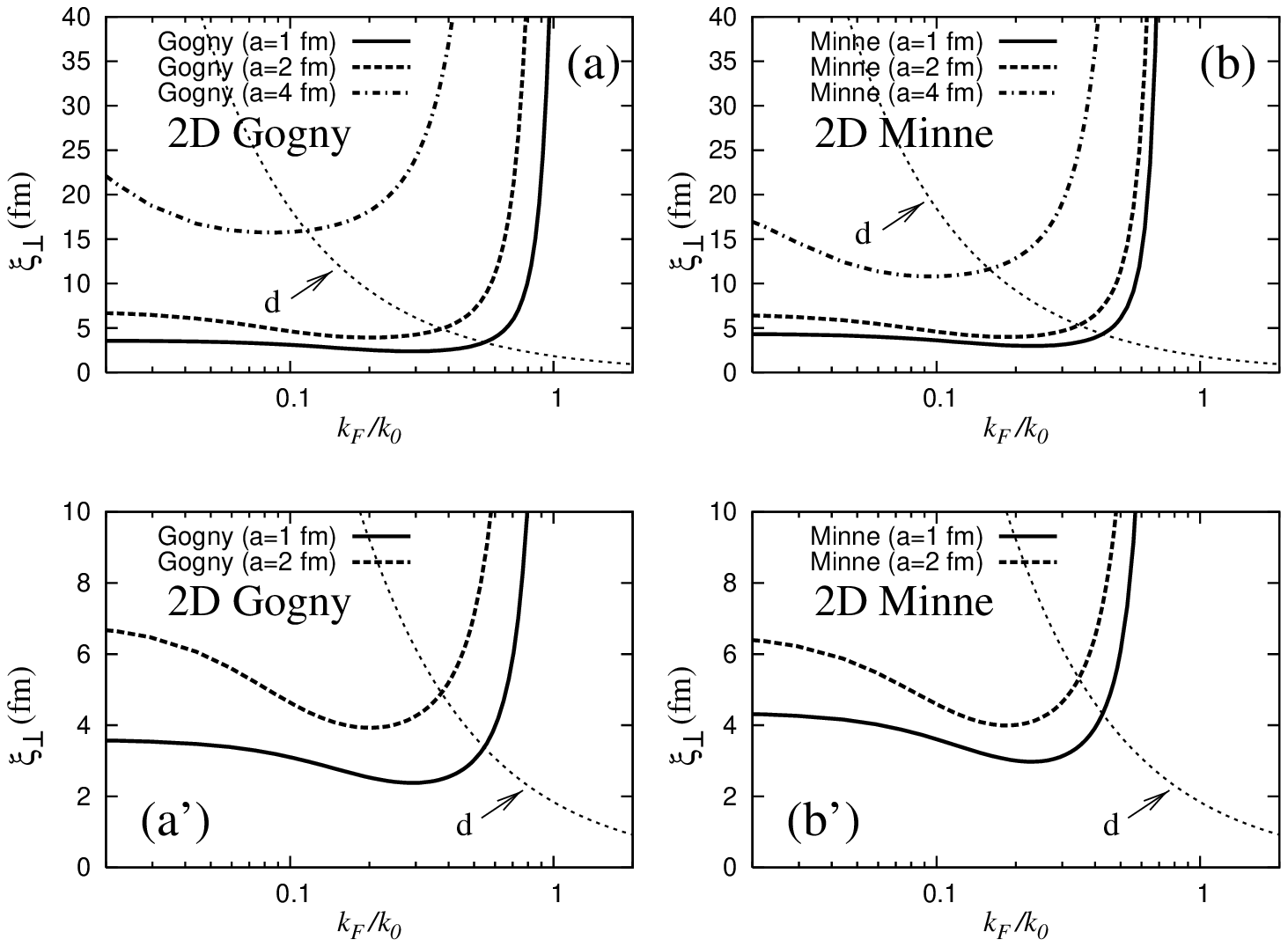}}
\vspace*{8pt}
\caption{(a)(b) Size $\xi_\perp$ of a 
Cooper pair $\Psi_{\rm pair}$ 
in quasi-2D neutron matter obtained with the Gogny D1S and Minnesota forces.
The calculated values are plotted as a function of 
$k_F/k_0$ ($k_0=1.36$ fm$^{-1}$). 
The average inter-neutron distance $d= \rho^{-1/2}$ is plotted 
with the dotted lines. (a')(b') Same as (a)(b) but scaled 
up for the vertical axis. 
\label{fig:distance}
}
\end{figure}

\begin{figure}[th]
\epsfxsize=6 cm
\centerline{\epsffile{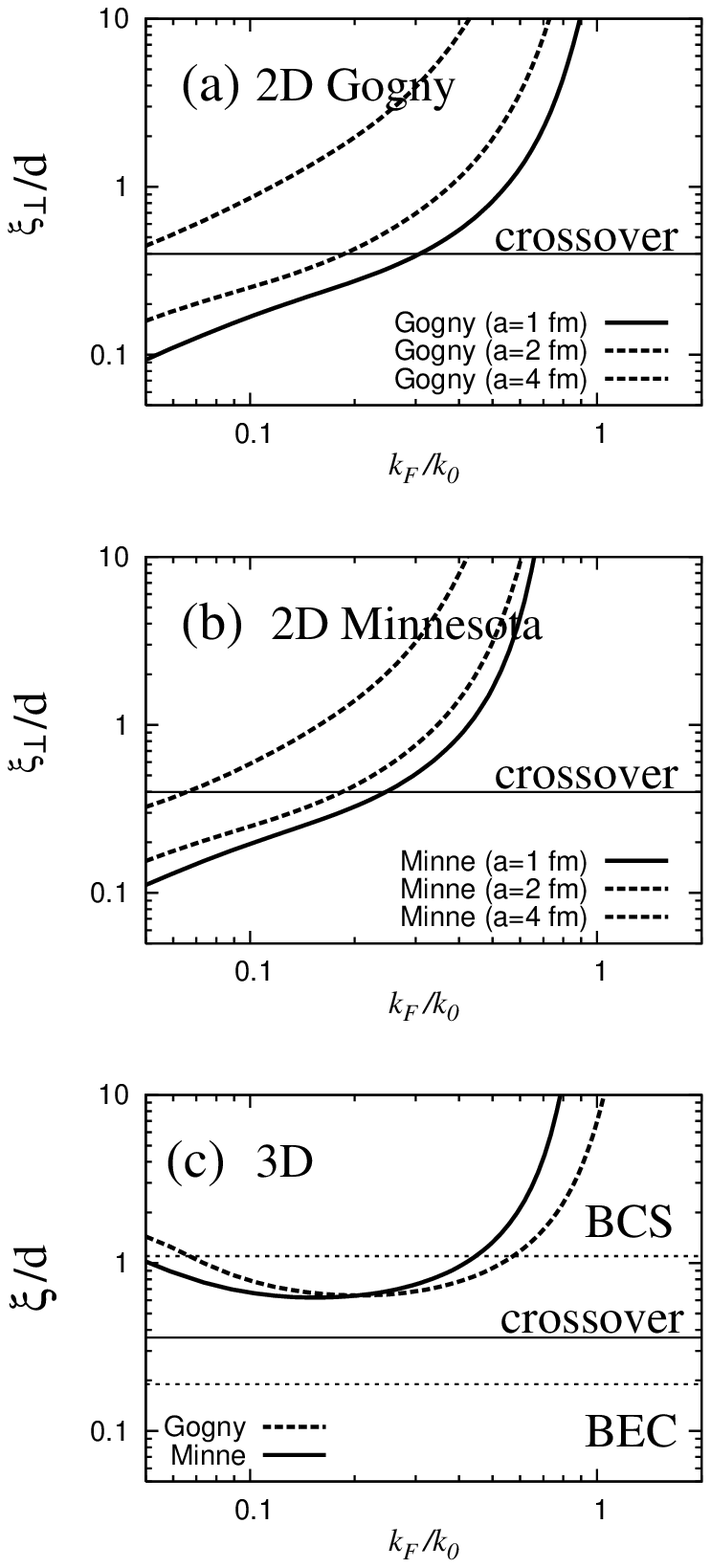}}
\vspace*{8pt}
\caption{Ratio of the $nn$-Cooper pair size
to the average inter-neutron distance $d$ as a function of 
$k_F/k_0$ ($k_0=1.36$ fm$^{-1}$).
(a)(b) Ratio $\xi_\perp/d$ ($d=\rho^{-1/2})$ 
in quasi-2D neutron matter obtained with the 
Gogny D1S and Minnesota forces.
The $k_F\xi_\perp=1$ line for a typical value of the 
BEC-BCS crossover \cite{randeria89,lotkev01,pistolesi96} in 2D is shown by solid thin lines.
(c) Ratio $\xi/d$ ($d=\rho^{-1/3})$ in 3D neutron matter 
with the Gogny D1S and Minnesota forces.
The short-dashed lines indicate the boundaries for 
the weak coupling BCS region $\xi/d> 1.1$, 
the crossover region $0.19 \le \xi/d \le 1.1$
and the strong coupling BEC region $\xi/d <0.19$ 
in the regularized 
model \cite{Matsuo:2005vf,Engelbrecht:1997zz,Papenbrock:1998wb}.
Thin solid line is $\xi/d =0.36$ for the unitary limit. 
\label{fig:rmsr-r}
}
\end{figure}

The pair size $\xi_\perp$ in
quasi-2D neutron matter is plotted as a function of $k_F/k_0$ 
in Fig.~\ref{fig:distance}. 
In the $k_F\rightarrow 0$ limit, 
the pair size $\xi_\perp$ equals  the size of 
the two-neutron bound state ($\xi^{nn}_\perp$ in (\ref{eq:xinn})).
With the increase of $k_F$, the pair size reduces first 
before it becomes large again with further increase of 
$k_F$. 
Thus the minimum pair size $\xi_\perp$ is found to be 
smaller than the size $\xi^{nn}_\perp$ of the isolated two-neutron bound state 
in the quasi-2D system. This size reduction of the 
$nn$-Cooper pair indicates the 
enhancement of the dineutron correlation at finite low $k_F$, i.e. due to 
the existence of a Fermi sea.  
It should be pointed out that the pair size significantly
depends also on the width $a$ of the slab.

A striking difference of the pair size between the quasi-2D and 
3D neutron matter is that
the pair size at $k_F\rightarrow 0$ limit 
is finite in the quasi-2D neutron matter 
while it becomes infinite in the 3D neutron matter. 
In the 3D neutron matter, the Cooper pair size $\xi$ 
has a minimum around $k_F/k_0 = 0.5$, and it rapidly
increases as $k_F$ goes 0.
Interestingly, the size shrinking of the Cooper pair
in quasi-2D neutron matter has a close analogy 
with that of the deuteron in 3D symmetric matter 
predicted by Lombardo and Schuck \cite{Lombardo:2000tm}.

As mentioned above, the spatial correlation of Cooper 
pairs enhances at finite low $k_F$ in  quasi-2D neutron matter with a 
small width $a$, where the BCS-BEC crossover phenomena is expected.
To discuss pairing properties from the point of view of 
the BCS-BEC crossover, it is useful to compare 
the pair size $\xi_\perp$ with 
the average inter-neutron distance $d\equiv \rho^{-1/2}$
as discussed 
in the works for 3D neutron matter \cite{Matsuo:2005vf,Margueron:2007uk}.
The ratio $\xi_\perp/d$ in quasi-2D neutron matter 
is shown in Fig.~\ref{fig:rmsr-r} as well as
$\xi/d$ for the 3D neutron matter.
Strictly speaking, because 
the transition from the weak coupling BCS phase to 
strong coupling BEC is known to be a crossover \cite{Nozieres:1985zz},
there is no critical boundary of the transition. However, 
we consider that BCS-BEC crossover features arise    
when the Cooper pair size is smaller than the average inter-neutron 
distance by adopting the same criterion as in 
Refs.~\cite{Matsuo:2005vf,Margueron:2007uk}.
With this criterion, the BCS-BEC crossover phenomena appears 
at $k_F/k_0< 0.4\sim 0.5$ in quasi-2D neutron matter with $a=1$ fm, and
$k_F/k_0 < 0.3\sim 0.4$ for $a=2$ fm, while in 3D neutron matter
it is seen at $k_F/k_0< 0.4 \sim 0.5$. 
In the quasi-2D neutron matter, the ratio $\xi_\perp/d$ monotonically deceases 
as $k_F$ becomes small, and finally the system goes to the
strong coupling BEC limit.
 
\begin{figure}[th]
\epsfxsize=16 cm
\centerline{\epsffile{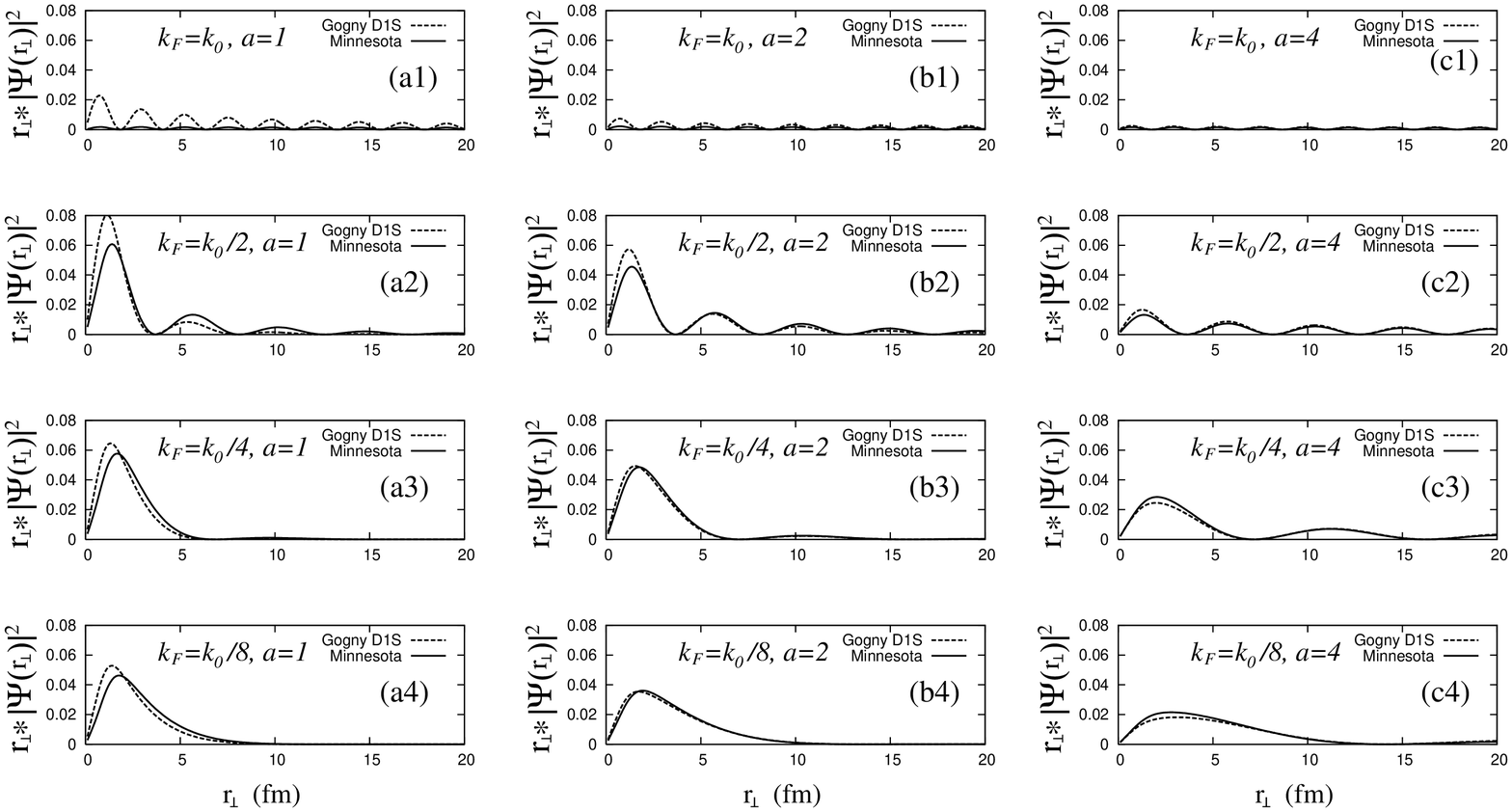}}
\vspace*{8pt}
\caption{Probability density 
$r_\perp |\Psi_{\rm pair}(r_\perp)|^2$ of a $nn$-Cooper pair
in the quasi-2D neutron matter
as a function of the relative distance $r_\perp$.
(a1)(a2)(a3)(a4) Probability density in the slab with $a=1$ fm
at $k_F=k_0$, $k_F=k_0/2$, $k_F=k_0/4$, $k_F=k_0/8$. 
(b1)(b2)(b3)(b4) The density in the slab with $a=2$ fm.
(c1)(c2)(c3)(c4) The density in the slab with $a=4$ fm.
\label{fig:wf-2d-rpp}
}
\end{figure}

\begin{figure}[th]
\epsfxsize=12 cm
\centerline{\epsffile{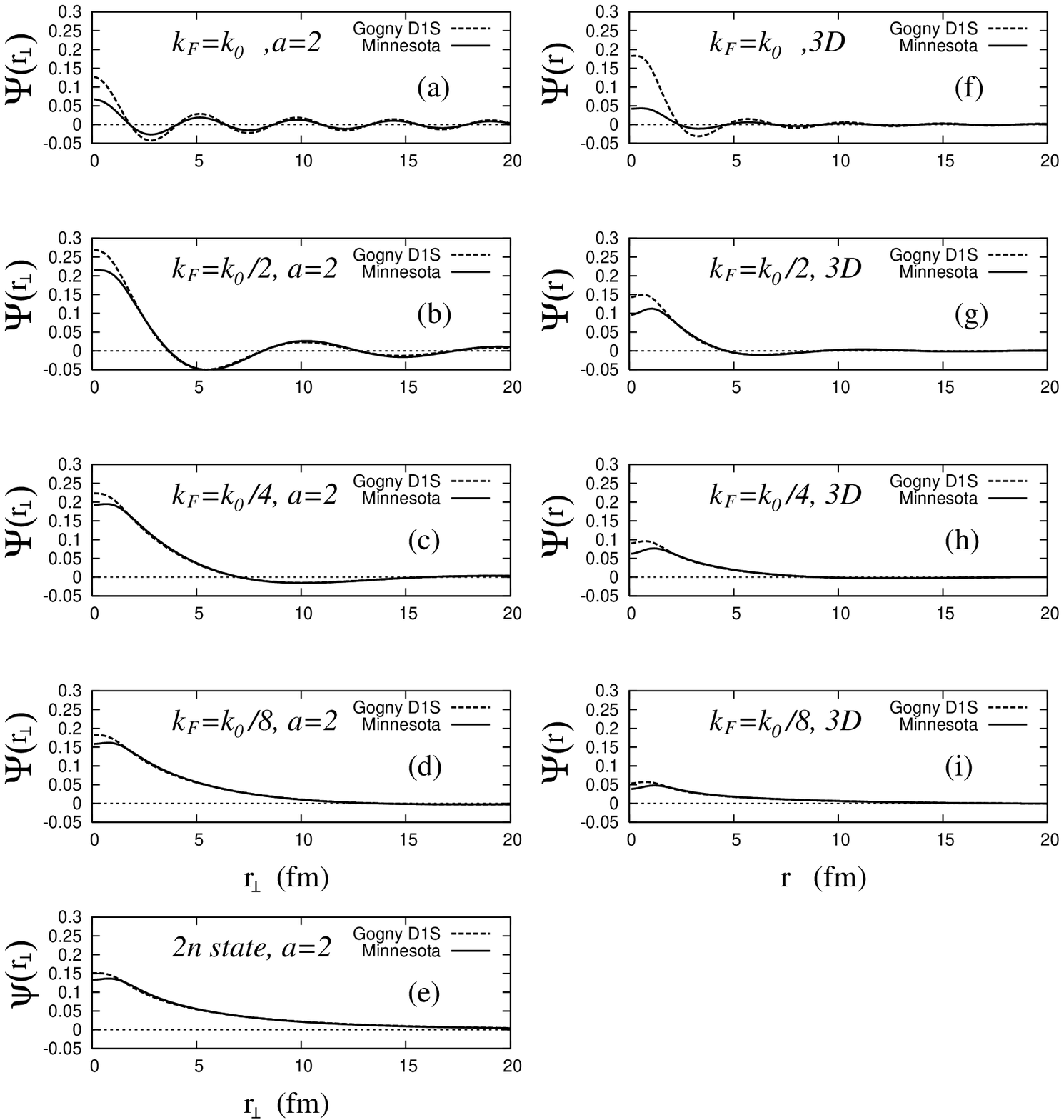}}
\vspace*{8pt}
\caption{(a)(b)(c)(d) Pair wave function 
$\Psi_{\rm pair}$ of a $nn$-Cooper pair
in the quasi-2D neutron matter with $a=2$ fm as a function of $r_\perp$, 
and (f)(g)(h)(i) that in 3D neutron matter as a function of $r$.
(e) Two-neutron wave function $\psi^{nn}(r_\perp)$ of 
the bound state in the quasi-2D system with $a=2$ fm.
\label{fig:wf-phi}
}
\end{figure}

To reveal the features of BCS-BEC crossover
it is also
useful to analyze the spatial structure of the $nn$-Cooper 
pair wave function\footnote{
We are aware that the notion 'Cooper pair wave function' has, recently, become 
a subject of debate (see, e.g., G.G. Dussel, S. Pittel, 
J. Dukelsky, P. Sarriguren, Phys. Rev. 
C {\bf 76}, 011302 (2007) and references in there). 
We do not want to enter this discussion here and 
stay with the traditional jargon.}
as done in Refs.~\cite{Matsuo:2005vf,Margueron:2007uk}.
Figures \ref{fig:wf-2d-rpp} 
and \ref{fig:wf-phi} show the
probability density $r_\perp|\Psi_{\rm pair}(r_\perp)|^2$ 
and pair wave function
$\Psi_{\rm pair}(r_\perp)$ 
of the Cooper pair in the quasi-2D neutron
matter. 
Let us discuss the results for $a=2$ fm.
At low $k_F$ such as $k_F=k_0/8$ and $k_F=k_0/4$
(Fig.~\ref{fig:wf-2d-rpp}(b4) and (b3)), 
the spatial correlation of the Cooper pair 
is strong and it indicates a BEC-like dineutron pair.
From $k_F=k_0/8$ to $k_F=k_0/4$, 
the first peak becomes narrow
due to the Pauli principle, stemming from the other Cooper pairs. 
Consequently, 
the size shrinking of the dineutron occurs at a finite low
$k_F$, as mentioned.
With further increase of $k_F$, the nodal structure appears 
at $k_F=k_0/2$ which indicates the transition  from the 
BCS-BEC crossover to the BCS regime, and at $k_F=k_0$ the short-range 
correlation disappears as in the weak coupling BCS phase.

The pair wave functions in quasi-2D neutron matter 
with $a=2$ fm are shown in Fig.~\ref{fig:wf-phi}.
We also show the two-neutron wave function $\psi^{nn}(r_\perp)$
for the dineutron bound state in the quasi-2D system
and $\Psi_{\rm pair}(r)$ in 3D neutron matter.
In the $k_F\rightarrow 0$ limit, the pair wave function 
$\Psi_{\rm pair}(r_\perp)$
becomes equal to the isolated two-neutron wave function 
$\psi^{nn}(r_\perp)$ of the dineutron state
in case of quasi-2D neutron matter, while $\Psi_{\rm pair}(r)$ 
in 3D neutron matter becomes infinitely broad because there is 
no two-neutron bound state in a 3D system.

\section{Discussion}\label{sec:discussion}

\subsection{Possible relation to neutron pair correlations in neutron skins}

In this section we tentatively associate the strong pairing correlations 
of valence neutrons in neutron-rich nuclei with the ones in a 2D system. 
It has recently become evident that in neutron-rich or neutron halo 
nuclei pairing correlations may show a strong maximum in the surface. This 
is for instance evidenced by the fact that the size of the Cooper 
pairs has a sharp minimum in the surface. This feature has been shown 
to hold true in the case of a single Cooper pair \cite{Hagino:2006ib} 
as well as 
in cases of many Cooper pairs \cite{Pillet:2007hb}.
Though for the moment, it cannot be excluded that the pronounced small Cooper 
pair sizes in the nuclear surface is due to a somewhat trivial, i.e. 
geometrical effect\cite{Sandulescu09}, we here make the hypothesis that this is due to a real 
strong correlation effect, due to the presence of a surface and hence of a 2D 
influence.

The valence nucleons may be confined in a surface shell for different 
reasons. 
For example, if valence neutrons are, in a heavy nucleus, distributed 
around a high lying principal shell, the corresponding density distribution, 
for instance if the shell does not contain an $S$ wave, may be concentrated in 
the surface region.
Even though low angular momentum orbits with radial nodes have some
strength distribution in the inner region, their probability amplitudes 
may show peak structure 
in the surface, and therefore, the pairing is expected to be 
induced by the surface neutrons due to the possible  
mixing of many states  having significant overlap of their radial part 
in the surface region. 
 
On the other hand it is also well known 
that finite size {\it enhances} pairing. 
This is already documented by the fact 
that the nuclear gap varies on average like $12/\sqrt{A}$ MeV, that is the 
presence 
of a surface certainly plays a role. This finite size enhancement is a 
quite subtle effect as theoretical investigations show \cite{farine03} and 
implies several competing effects. How much the scenario is influenced by 
quasi 2D features is still an open question but an interesting 
and suggestive hypothesis which we wanted to explore in this work. So one may 
consider that the strongly correlated Cooper pairs with about a rms 
radius of 2 fm live in a surface sheet about $1-2$ fm across. In section 
\ref{sec:formulation} we mimicked the transverse motion of the neutrons 
by a simple frozen 
Gaussian $\phi_0(z)$. The motion perpendicular to $z$ is considered as 
planar. This is, of course, a highly idealized representation of the 
situation as mentioned several times.
It may nevertheless capture some 2D features which possibly 
underly the 
strong pairing correlations in the neutron skin of neutron-rich nuclei.

In the previous analysis of pair size in quasi 2D neutron matter in 
section \ref{sec:results}, the BCS-BEC crossover 
may occur only in dilute 2D matter for 
densities $\rho \le 0.05-0.07$ fm$^{-2}$. Let us estimate how many small sized 
Cooper pairs can condensate in an area $4\pi R^2$ of the nuclear surface. 
For the dineutron condensation, more than one pair should exist at least.
From the condition $\rho \le 0.05-0.07$ fm$^{-2}$, 
two dineutrons can condensate 
in the surface area for $R\ge 2-3$ fm, and four dineutrons 
for $R\ge 3-4$ fm.

Recently, two-neutron correlations in neutron-skin nuclei 
have been investigated in light nuclei
such as $^8$He and 
$^{18}$C \cite{KanadaEn'yo:2007ie,Itagaki:2008zz,Hagino:2008vm}
as well as medium-heavy nuclei \cite{Matsuo:2004pr,Pillet:2007hb}.
$^8$He is considered to be a neutron-skin nucleus 
having four valence neutrons surrounding an $\alpha$ core, 
and therefore this nucleus is one of the simple examples to discuss 
neutron pair correlations in neutron skins.
In Refs.~\cite{KanadaEn'yo:2007ie,Itagaki:2008zz,Hagino:2008vm},
it was argued that the $^8$He ground state has 
spin-zero two-neutron correlations.
From the above-mentioned discussion of quasi-2D neutron matter, 
the correlation in the ground state is expected to be of BCS-type 
rather than in the BCS-BEC crossover region,  
because the valence neutron distribution 
concentrates in the surface with a distance $R\le 2$ fm, 
which is smaller than the BCS-BEC crossover 
condition $R\ge 2-3$ fm for the
condensation of two dineutrons.
Instead of the ground state, the possible existence of an 
excited state $^8$He($0^+_2$) with a dilute gas-like state
of dineutrons was suggested by one of the 
authors \cite{KanadaEn'yo:2007ie}. 
According to the theoretical results of Ref.~\cite{KanadaEn'yo:2007ie},
the dineutrons exist around the region with $R=4-5$ fm 
which satisfies the above-mentioned 
condition for the BCS-BEC crossover.
Therefore, this state could have BEC-like dineutrons 
at the surface.

In medium-heavy nuclei, the spatial structure of 
spin-zero neutron pairing at the surface was investigated 
based on HFB calculations in Refs.~\cite{Matsuo:2004pr,Pillet:2007hb}.
To discuss two-neutron correlations in the neutron
skin in connection with quasi-2D neutron matter, 
one should analyze the radial and angular behaviour
separately and focus on the angular correlations (opening angle) of
neutron pairing, as well as on the size of the Cooper pairs.
The transition from BCS phase to BCS-BEC crossover is 
characterized by the change of the 
pair wave function from a weak-coupling nodal structure 
to an enhanced single bound state like peak. 
If such a structure change is found
in the angular behaviour it may be an indication of the BCS-BEC
crossover, while a clear evidence may not be seen 
in the radial structure because it might be affected 
by Pauli effect from constituent neutrons of the core.
The spatial structure of spin-zero neutron pairing 
is displayed in Ref.~\cite{Matsuo:2004pr}, 
which shows
that the nodal structure along the angular direction 
is seen inside from the surface and 
it disappears on the outer side of  the surface. 
More detailed analysis of the angular behaviour is requested.
On the other hand, a good measure of the correlations also is the size of the 
Cooper pairs. Their rms radius seems systematically very small in the outer 
part of the nuclear surface \cite{Pillet:2007hb}, that is much smaller than 
the smallest value in 
homogeneous 3D matter. Whether this also always goes along with a maximum in 
the pair correlations at the same place in a finite nucleus, remains to be 
seen.

Let us also comment on other effects omitted in the present model of 
infinite quasi-2D neutron matter.
At the surface of a realistic finite nuclear system, 
valence neutrons feel the
centrifugal force and the spin-orbit force 
which may work so as to break $nn$-Cooper pairs. 
Periodicity condition
on the spherical surface is another effect to be considered.

\subsection{Relation to two-neutron halo nuclei}

Two-neutron correlations have extensively been investigated
in two-neutron halo nuclei by means of 
three-body models of two neutrons and a core 
\cite{Hansen:1987mc,Bertsch:1991zz,Zhukov:1993aw,Esbensen:1997zz,ikeda02,
aoyama01,Myo:2002wq,Myo:2003bh,Hagino:2005we,myo07,baye97,arai01,Descouvemont:2005rc}.
Recently, it was suggested that the 
internal two-neutron wave function analyzed as  
a function of the distance $R$ from the nuclear center
shows behaviour similar to the 
pair wave function of 3D neutron matter 
in the BCS-BEC crossover region \cite{Hagino:2006ib}.
In the analysis of Ref.~\cite{Hagino:2006ib},
the inter-neutron wave function has a
nodal structure at small $R$, and it shows
enhancement of the spatial correlations with increasing $R$.
Moreover, in a further analysis of the neutron pair in $^{11}$Li,  
size shrinking was found 
at the surface \cite{Hagino:2008wt}. It was found that there 
the pair is as small as
3 fm at $R\sim 3$ fm \cite{Hagino:2008wt}.
The formation of the small sized dineutron 
at the surface of $^{11}$Li 
may be understood with the picture of a quasi-2D neutron system. 
As shown in Ref.~\cite{Hagino:2005we} with a three-body model,  
the spin-zero component of two neutrons shows a strong angular 
correlation and its distribution concentrates at the surface with 
$R=3\sim 4$ fm.
The situation corresponds to the quasi-2D neutron where the valence neutrons
are confined at the surface region and the neutron pair 
is formed due to 
the angular correlation. It may be interpreted as 
the reduction of the dimension from three to two, being 
crucial for the small-sized neutron 
pair in two-neutron halo nuclei.
At the small $R$ region, i.e. near the core, 
the size of the neutron pair increases and nodal structure 
appears in the relative wave function in the three-body model. 
This is
because the Pauli effect from the constituent neutrons of 
$^9$Li core raises the Fermi momentum of the 
valence neutrons. 
This behaviour is similar
to the weakening of the pair correlation in 
the BCS region at $k_F\ge k_0/2$ in the quasi-2D neutron matter.

In spite of the analogous behaviour 
of the neutron pair in $^{11}$Li
to that of the Cooper pair in the quasi-2D neutron matter,
one should keep in mind that two-neutron halo nuclei 
has only one pair of neutrons and 
the phenomena there is not regarded as 
the Cooper pair condensation 
in the BCS picture. Several investigations have, however, revealed that the 
features of a single Cooper pair and the ones of a Cooper pair embedded in 
several others are not drastically different.

\subsection{Two-proton correlation in quasi-2D systems}

It is also interesting to consider two-proton correlation 
in proton-rich nuclei 
in analogy to the two-neutron correlation in neutron-rich nuclei. 
While the energy of infinite proton matter diverges 
without charge neutrality, 
proton-rich matter such as proton halo or proton skin structure 
can exist in finite nuclei.
Moreover, since 
two-proton decays are allowed from some proton-rich nuclei 
such as $^6$Be,$^8$C,$^{12}$O and 
$^{16}$Ne \cite{Grigorenko:2001tb,Barker:2003sy,Mukha:2008ky}, 
more direct information of 
$T=1$ pair correlation could be experimentally obtained
than for the case of neutron-rich nuclei.

\begin{figure}[th]
\epsfxsize=6 cm
\centerline{\epsffile{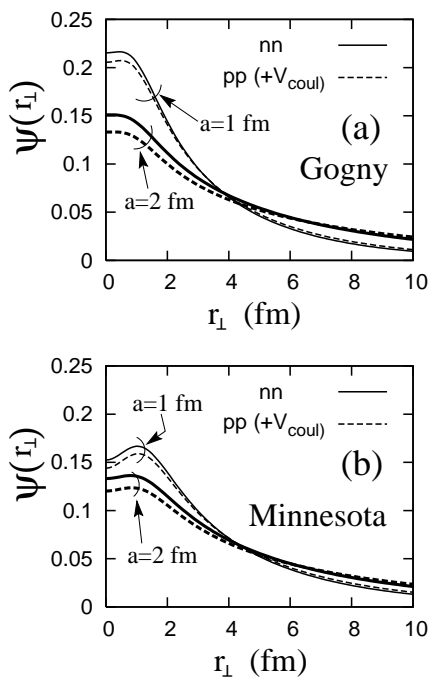}}
\vspace*{8pt}
\caption{Two-proton wave functions $\psi^{2p}(r_\perp)$ 
in quasi-2D systems with the width parameter
$a=1$ and $a=2$ fm calculated with (a) the Gogny force and
(b) Minnesota force complemented with Coulomb force ($V_{\rm Coul}$).
Two-neutron wave functions $\psi^{nn}(r_\perp)$ (without $V_{\rm Coul}$) 
are also shown for
comparison.
\label{fig:wf-2p}
}
\end{figure}

In the same way as neutron systems, 
we calculate a wave function $\psi^{2n}(r_\perp)$ for the
 two-proton bound state in 
quasi-2D systems. Coulomb force ($V_{\rm Coul}(r)$) is added to the two-body
effective nuclear forces.
In Fig.~\ref{fig:wf-2p}, the obtained two-proton wave functions
in the $^1S$-wave channel 
are shown in comparison with
those of the dineutron state.
Because of the Coulomb force, the spatial correlation 
of the $T=1$ pair is slightly more suppressed in the two-proton system
than in the two-neutron system. However, 
the wave functions for $a=1$ fm and $a=2$ fm still 
have a pronounced peak. It suggests that the spatial correlation of the 
$T=1$ pair also is present in proton systems.

\section{Summary and Outlook}\label{sec:summary}

Singlet $S$ ($^1S$) $nn$ correlations 
in quasi-2D neutron systems 
were studied in a simplified model of 
neutrons confined in a slab with a certain thickness.
The finite-range effective nuclear forces, 
the Gogny D1S and Minnesota forces, were adopted for the study.
The strength of the two-body interaction 
in the quasi-2D systems depends on the width $a$ of the slab.
Namely, the interaction is stronger for thinner slabs than for
wider ones.
When the width of a slab 
is small enough, e.g. around $a=1$ fm,  
two neutrons form a tightly bound dineutron with a size as small as
the deuteron size.

The behaviour of the $nn$-Cooper pair in the 
quasi-2D neutron matter was investigated
by means of the BCS theory. 
In  quasi-2D neutron matter, 
the transition region of the BCS-BEC crossover 
appears at $k_F/k_0< 0.4\sim 0.5$ and for $a\sim 1$ fm.
The size shrinking of the Cooper pair occurs 
at finite low $k_F$ because of the Pauli effect.
It is also important to note that, as the width $a$ of the slab decreases, 
the pair correlation in 2D is enhanced and the Cooper pair size 
becomes small because of the stronger attraction 
in the thinner slab. 
This is one of the characteristics different from 3D.
In particular, the reduction of the dimension from 3D to 2D is 
crucial to form the strongly correlated neutron pair. The chemical potential 
becomes negative for $k_F/{k_0}<0.2$ and, therefore, there is true two-neutron 
binding in quasi-2D neutron matter and a genuine BEC regime is born out.

The relation of the quasi-2D neutron matter 
to finite neutron-rich nuclei was also discussed.
Several studies of the past have revealed that the size of Cooper pairs 
changes very substantially from inside to outside of 
a nucleus \cite{Hagino:2008wt,Pillet:2007hb}.
The subject 
has been taken up again more recently and the radius dependence of a 
neutron-neutron ($nn$) Cooper pair has
been investigated more systematically  \cite{Hagino:2008wt,Pillet:2007hb}. 
It thereby turned out that the old believe that the size of a 
Cooper pair is of the order of the diameter of a large nucleus must be 
revisited when considered locally, i.e., as a function of the nuclear radius. 
Namely, detailed BCS calculations with the D1S Gogny force have shown that 
the rms radius of an $nn$-Cooper pair, though indeed 
large in the interior, 
becomes very small in the outer surface of a nucleus, where the 
density is low, before expanding again when leaving the nucleus. 
The minimum extension of a Cooper 
pair attains a size as small as 2 fm, that is the same 
as the size of a deuteron, a 
bound state. Although the variation of the pair radius 
as a function of the density 
qualitatively follows the one of infinite matter, 
a factor of two reduction of the pair diameter with 
respect to the homogeneous case is a quite surprisingly strong effect 
which calls for an 
explanation. 

We considered  this pair size shrinking in the surface of 
neutron-rich nuclei with the present calculations of 
quasi-2D neutron systems, based on the hypothesis that the valence 
neutrons in a neutron-rich nucleus live in radial direction in a surface 
sheet about $1-2$ fm across because the wave functions of the active shell 
are essentially 
concentrated in the surface while the nucleons are free to move 
within the sheet. The idea behind this scenario is that the neutrons in the 
surface sheet are essentially constrained to a two-dimensional world which
could give a certain clue to the smallness of the Cooper pairs, since in two 
dimensions even an isolated $nn$ pair forms a bound state. Pursuing this idea, 
we further idealized the situation assuming a planar slab of neutrons where 
for simplicity the motion perpendicular to the slab is frozen into a Gaussian 
wave packet in order to simulate the surface concentration of the valence 
neutrons in radial direction. Freezing the perpendicular motion means that the 
neutrons are two-dimensional in the direction of the slab. The thickness of 
the slab only influences the strength of the pairing force, 
as already mentioned above. 
We, indeed, find that for 
reasonable slab thicknesses of $a \sim 1$ fm, the pair radius is of the 
order of 2$-$3 fm in agreement with realistic 3D HFB 
calculations \cite{Pillet:2007hb}.
We are, of course, aware of the fact that our model may be strongly 
oversimplified and further studies are required 
in order to ascertain whether a 2D 
aspect indeed helps to interpret the occurrence of astoundingly small 
Cooper pairs in the 
nuclear surface. One also should keep in mind that a local picture may be a 
little too simple in the sense that the pair size certainly fluctuates a lot.
Nevertheless, a classical picture often grasps a good deal of reality, even 
in so tiny quantum object as nuclei. 
Neutron pairs much smaller than the dimension of a nucleus may approximately 
be considered as bosons and a strong coupling Bose-Einstein condensation 
scenario of $nn$ pairs may be realized in the surface of neutron-rich nuclei.
Bound state like Cooper pairs in the 
nuclear surface may also allow to reinterpret neutron transfer reactions into 
superfluid nuclei which are known to have a strongly enhanced cross 
section \cite{oertzen01}. 

2D aspects may also be relevant in the Lasagne 
phase in neutron star crust.
In practical application to such systems, 
one should extend the model by taking
into account finite temperature and also excitation 
in the $z$-direction for higher neutron density.

BCS and BEC problems of two-dimensional fermion systems
have been discussed also 
in the field of condensed matter 
physics in relation to electron systems and 
atomic systems \cite{randeria89,lotkev01}. 
Phenomena of BEC and BCS-BEC crossover 
in quasi-2D systems of trapped gases have been 
discussed in recent works \cite{petrov00,he08,Zhang08}, and 
it has been suggested that the condensation in a
quasi-2D gas is sensitive to the frequency $\omega$ 
of the confinement 
in the ``frozen'' direction.
This feature may have an analogy to the present results 
for the quasi-2D neutron matter.

\section*{Acknowledgements}
The authors would like to thank K.~Hagino, H.~Horiuchi, N.~Pillet, G.~R\"opke, H.~Sagawa and
N.~Sandulescu 
for valuable discussions on nuclear pairing and Cooper pairs in the 
nuclear surface.
Fruitful discussions during
the YIPQS international molecule workshop held at Yukawa Institute for 
Theoretical Physics (YITP) in October 2008 
were useful to complete this work. 
The computational calculations of this work were performed by 
supercomputers at YITP and also by Supercomputer Projects 
of High Energy Accelerator Research Organization (KEK).
This work was supported by Grant-in-Aid for Scientific Research from Japan Society for the Promotion of Science (JSPS).
It is also supported by 
the Grant-in-Aid for the Global COE Program "The Next Generation of Physics, 
Spun from Universality and Emergence" from the Ministry of Education, Culture, Sports, Science and Technology (MEXT) of Japan. 

\end{document}